%% file: main.tex
\documentclass[runningheads]{llncs}
\usepackage{cite}
\usepackage{amsmath,amssymb,amsfonts}
\usepackage{graphicx}
\usepackage{textcomp}
\usepackage{xcolor}
\usepackage[hyphens]{url}
\usepackage{hyperref}
\usepackage{enumitem}

\def\BibTeX{{\rm B\kern-.05em{\sc i\kern-.025em b}\kern-.08em
    T\kern-.1667em\lower.7ex\hbox{E}\kern-.125emX}}

\usepackage{eurosym}
\usepackage{siunitx}
\usepackage{tikz}
\usetikzlibrary{arrows.meta,
		calc, chains,
		positioning,
		quotes 
		}
\usepackage{tikz-cd}
\usepackage[english]{babel}

\begin{document}

\title{Cybersecurity Competence for Organisations in Inner Scandinavia}
%\subtitle{}

\author{
    Simone Fischer-H{\"u}bner\inst{1}
    \orcidID{0000-0002-6938-4466},
	%\and
	Leonardo A.\ Martucci\inst{1}
	\orcidID{0000-0002-9980-3473}, 
    %\and
	Lejla Islami\inst{2}
	\orcidID{0000-0002-5717-8649},
    %and
    Ala Sarah Alaqra\inst{1}
	\orcidID{0000-0002-6509-3792} 
    and
    Farzaneh Karegar\inst{1}
    \orcidID{0000-0003-2823-3837} 
    }

\authorrunning{}
\authorrunning{Fischer-H{\"u}bner et al.}

\institute{
    Karlstad University, Karlstad, Sweden\\
	\email{\{simone.fischer-huebner, leonardo.martucci, as.alaqra, farzaneh.karegar\}@kau.se}
    \and
    Danmarks Tekniske Universitet, Lyngby, Denmark\\
    \email{lejis@dtu.dk}
    }

\maketitle

\input{src/abstract}
\input{src/intro}
\input{src/background}

\input{src/Method}
\input{src/results-survey}
\input{src/results-interviews}

\input{src/discussion}
\input{src/conclusions}

\section*{Acknowledgements} This study was co-funded by the Cross Border Cyber Capacity (CBCC) project, an Interreg Sverige-Norge cross-border cooperation, innovative environments project, which is co-funded by the European Union.

\bibliographystyle{splncs04}
%\bibliography{main}

\input{src/appendices}

\end{document}

%% file: src/abstract.tex
\begin{abstract}
A rapidly growing number of cybersecurity threats and incidents demands that Swedish organisations increase their efforts to improve their cybersecurity capacities. This paper presents results from interviews and a prior survey with key representatives from enterprises and public sector organisations in the Swedish region of V{\"a}rmland in Inner Scandinavia, examining their cybersecurity readiness and needs for education and competence development. We discuss the generalizability of our findings and the extent to which they may be specific to Sweden and Värmland, and we conclude by proposing efforts to strengthen cybersecurity competences in Inner Scandinavia.

\end{abstract}

%% file: src/intro.tex
\section{Introduction}
\label{s-intro}
Cybersecurity attacks and incidents with a drastic impact on organisations are steadily increasing worldwide. In Inner Scandinavia, especially in the Swedish region of V\"{a}rmland, recent incidents included distributed denial of service attacks 
 against the supermarket chain Coop in December 2023, and the IT cloud-hosted services provided by the Finnish company Tietoevry, which serves public institutions, in 
 January 2024.\textsuperscript{1,2} 
 While Sweden ranks fourth globally in digitalisation, % in the world, 
 it lags in cybersecurity readiness, placing 16$^{th}$ in the National Cybersecurity Index.\textsuperscript{3} This gap between digitalisation and cybersecurity readiness levels makes Sweden especially vulnerable to attacks and calls to improve readiness and competence. Sweden's new NATO membership further increases the need to invest in defence and strengthen cybersecurity capacity, knowledge, and skills, as outlined in Sweden's new Cybersecurity Strategy for 2025--2029.
Advanced competence is also required
to understand and prepare for new cybersecurity rules and regulations, including the Network and Information Security Directive 2 (NIS2), the Cyber Resilience Act (CRA), the Digital Operational Resilience Act (DORA), 
and the Artificial Intelligence Act. 
The high number (``tsunami'') of such regulations challenges European organisations in meeting compliance requirements for risk analyses and implementing adequate security measures.

\setcounter{footnote}{3}
\footnotetext[1]{\url{https://pressrum.coop.se/coop-varmland-information-om-cyberangrepp/}}
\footnotetext[2]{\url{https://www.tietoevry.com/en/newsroom/all-news-and-releases/other-news/2024/01/ransomware-attack-in-sweden-update/}}
\footnotetext[3]{\url{https://ncsi.ega.ee/ncsi-index/?order=rank&archive=1}}

On the other hand, the V{\"a}rmland region in Inner Scandinavia is strategically 
%labeled as a ``safe'' region by the Swedish Civil Contingencies Agency (MSB), especially the Värmland is 
considered a ``safe'' region in Sweden due to its inland location. The region has a robust network of fibre-optic cable connections between its capital and Stockholm. It hosts the Swedish Civil Contingency Agency (MSB) headquarters, and accommodates Telia's (the largest Swedish Telecom provider) security operations centre. 
Neighbouring Innlandet, Norway, has developed as a Norwegian cybersecurity hub, hosting companies and a research centre.
The Interreg Sweden--Norway project [XXX]~\footnote{Anonymised for blind review} aims to strengthen V\"{a}rmland and Innlandet in cyber-- and societal security competence, competitiveness and readiness through cross-border cooperation and cybersecurity training and education. 

The goal of our work %in the [XXX] project 
is to examine the status of cybersecurity competence and readiness, along with challenges, requirements, and needs for cybersecurity and competence development measures for different types of organisations in Inner Scandinavia.
%, and more specifically in the two border-regions Värmland and Innerlandet Norway that participate in the CBCC project. 
As a first step, the project partners conducted surveys and interviews in Värmland and in Innlandet to examine these aspects in each region separately. The results inform discussions on enhancing cybersecurity competence across borders in Inner Scandinavia through cooperation.
To this end, the article addresses the following research questions for different types of organisations in Värmland:
\begin{enumerate}[label={RQ\arabic*}:, leftmargin=*]
    \item What is the current state of cybersecurity and
preparedness for these organisations?
    \item What are the cybersecurity education and competence development challenges for the organisations?
    \item What are their main interests and requirements regarding cybersecurity education and competence development? 
    %(regarding topics, target groups, and course forms)
%\item[RQ4:] Are there similarities and differences between organisations in Värmland and Innerlandet Norway?
\end{enumerate}

In this paper, we report and discuss the results from the two user studies that we conducted in V{\"a}rmland: (\emph{1}) an online survey (N=19) and (\emph{2}) semi-structured interviews (N=6) with key stakeholders from different types of private and public organisations, including small to medium-sized enterprises (SMEs) and large-sized companies and public sector organisations.
% While the survey was the starting point for our study, the focus of this article will be on reporting and discussing our explorative and qualitative research results from the interviews that provided us with deeper insights and understanding. 
While the survey informed our study, this article focuses on reporting and discussing insights from the exploratory qualitative interviews. This article makes the following contributions:
\begin{itemize}
    \item Detailed insights and analysis of the cybersecurity readiness, challenges, and needs for education and competence development among public and private organisations in Värmland in Inner Scandinavia.
    %, including small-medium-sized and large-sized companies, in Värmland in Inner-Scandinavia.
    %, based on the conducted survey and semi-structured interviews with selected representatives of these organisations. 
    \item A discussion on whether the results are generally applicable or specific to Sweden and/or Värmland in Inner Scandinavia.
\end{itemize}

%% file: src/background.tex
\section{Background and Related Work}
\label{s-background}
\subsection{Requirements by new cybersecurity-related regulations}
The EU has released a broad range of cybersecurity laws to promote cyber resilience and protect its online economy and society.
NIS2 and CRA are important frameworks released as part of the EU's overarching Cybersecurity Strategy.

NIS2 replaced the NIS Directive from 2016 with rules to be implemented by national laws of the EU member states. NIS2 expands its scope to a broader range of public and private entities in the EU that provide ``essential or important'' services such as energy, health care, digital infrastructures, and food supply. It demands member states to strengthen cybersecurity by obliging critical-sector organisations to implement technical, operational and organisational cybersecurity risk management measures, and to report incidents. Specific governance rules require management bodies in these sectors to follow cybersecurity training and offer similar training to their employees regularly, so that they can, in particular, gain knowledge and skills for risk assessment and management.

Effective December 2027, the CRA complements NIS2 by mandating cybersecurity requirements for manufacturers and retailers of products with digital elements that can be connected to another device or network. It particularly requires mandatory risk assessments by manufacturers, security by design, security updates for the expected product lifecycle, and continuous vulnerability management.

Advanced risk assessment and management are also required by other new cybersecurity-related regulations, including DORA, which requires a risk management framework for financial entities, or the AI Act, which demands developers to conduct a fundamental rights impact assessment for high-risk AI systems. They come in addition to Data Protection Impact Assessments that may need to be conducted pursuant to the EU General Data Protection Regulation (GDPR). While the scopes of the laws and their risk assessment requirements are not the same, they still may overlap, and therefore organisations may also have to find approaches for aligning the different risk assessments for their organisations. 

\subsection{Cybersecurity competence mappings}
The European Cybersecurity Agency (ENISA) supports this effort by its map of European MSc cybersecurity programs and European Cybersecurity Skills Framework (ECSF)~\cite{ENISA2022ECSF}, which defines and categorises relevant roles and skills. 
 Cybersecurity readiness and shortcomings of Swedish private organisations have been measured, e.g., by the 2024 Cisco Cybersecurity Readiness Index based on a double-blind survey\footnote{\url{{https://newsroom.cisco.com/c/dam/r/newsroom/en/us/interactive/cybersecurity-readiness-index/documents/Cisco\_Cybersecurity\_Readiness\_SE.pdf}}}. MSB regularly surveys the public sector (with 41 questions) to assess systematic cybersecurity work and the needs for improvements~\cite{msb2024cybersakerhetskollen}.
 To our knowledge, this is the first study combining a questionnaire-based survey with semi-structured interviews of diverse stakeholders from the private and public sectors in Värmland / Inner Scandinavia. Thus, it offers in-depth insights into their cybersecurity situations, challenges, and needs for cybersecurity education and competence development.

%% file: src/Method.tex
\section{Methodology and participants}
\label{s-methods}
We first conducted an online survey on the SUNET Survey\&Report platform, which ran for four weeks in February 2024, that provided us with initial insights, which were further explored through interviews.
%We collected the data through an online survey on the SUNET Survey\&Report platform, which ran for four weeks in February 2024. 
The survey included three background/demographics (gender, role, organisational type), twelve main questions on educational needs and interests, and three concluding questions for general comments and willingness to join follow-up interviews. The main questions comprised six 5-point Likert items on the need for cybersecurity education for different groups (security experts, IT-knowledgeable users, lay users) and interest in courses on EU risk management compliance, plus four open-ended and two multiple-choice questions on preferred format and location. Survey participants (n=19) were recruited via a regional IT cluster mailing list: six from large enterprises, eleven from SMEs, and two from the public sector. Appendix~\ref{survey} lists the exact questions asked in the survey. 

Next, to gain deeper insights, we conducted a qualitative study with six semi-structured interviews with selected key representatives/stakeholders from various Värmland organisations between April 2024 and February 2025 (average duration: 40 minutes). The interview, in general, covered participants' current cybersecurity practices and readiness, perceived challenges and risks, and needs or preferences for cybersecurity training and competence development. The interview script is reported in Appendix~\ref{interviewguide}. Three participants were recruited from the survey follow-up list, and three via the authors’ professional networks. % Three participants from the online survey indicated that they would take part in a follow-up interview study and left their contact details also participated in the interviews. %those who indicated in the online survey that they would be available for a follow-up interview study and left their contact details. The other three participants were recruited via the authors' professional networks.
All interviewees were males with extensive IT experience. Their background and participant identifiers (P1-P6) are:
\begin{enumerate}[label={(P\arabic*})., leftmargin=*]
    \item The chief security officer from a municipality 
    %(ca.\,100,000 population).
    %
    \item The IT security coordinator from a higher education institute
    %university 
    %(ca.\,18,000 students).
    %
    \item The chief executive officer of an SME developing control systems and embedded software.
    \item A salesperson from an SME providing IT services and solutions.
    \item The security responsible for a mobile platform of a large company in the security and defence sector.
    \item The education coordinator of another large company in the security and defence sector.
\end{enumerate}
Our interview-based research was used as an explorative method for gaining a better understanding of the research challenges and deriving qualitative results. We stopped after six interviews, as 
%code saturation was reached, i.e. 
we reached a point
%in data analysis 
where no new themes or significant information emerged from the analysis.
%, indicating that we had likely captured  captured the range of experiences, perspectives, and issues relevant to answer our research questions.
Two authors participated in each interview, one was responsible for leading the interview and asking the questions, and one took notes. All interviews were audio-recorded, transcribed, and coded by at least two authors. The coding was then synchronised into a joint code book, from which themes were derived. This thematic analysis followed the methodology by Clarke and Braun~\cite{clarke2017thematic}.

Participation in the survey and interviews was voluntary with the consent of the participants, who were also informed that any questions could be left out. The survey study was conducted in compliance with the GDPR and the Swedish Ethical Review Act, and ethical approvals for both studies were obtained from the ethics advisor at the authors' faculty.

%% file: src/results-survey.tex
\section{Survey and Interview Results}
\subsection{Survey Results}
\label{ss-survey-results}
%
%We first conducted an online survey, which is briefly described in this section. It provided us with initial insights, which were further investigated through interviews.

%\subsection{Methodology and participants}
%

%\subsection{Results}
%\label{ss-survey-results}
Although the limited number of responses from organisational representatives does not allow for fully representative conclusions for all organisations in Värmland, they offered initial insights into educational interests and needs These insights were later explored further in the follow-up interview study, revealing findings such as: %which we could then follow-up in the subsequent interview study, regarding educational interests and needs for different types of organisations, including the following findings: 
\begin{itemize}
    \item 
    Most participants from \textbf{large companies} (4 of the 6) indicated a need for advanced education of cybersecurity experts, including penetration testing training. They were less interested in other courses, including anti-phishing training, for other target groups. The majority (5) of them indicated that their organisations were well or somewhat prepared to meet the new EU cybersecurity risk management requirements, and none expressed interest in related courses.
   % : Most large companies have  an interest in education of cybersecurity experts, incl. penetration testing – less interest in other type of courses. It can be assumed  that they have already phishing security/ awareness training in place and thus less interest in such type of education.
%* They seem to be better prepared for the new laws and risk management requirements (at least they state so) and therefore not interested in such type of courses.
    \item In contrast, most SME participants (8 of 11) expressed interest in courses to prepare for EU cybersecurity risk management requirements and basic training (e.g., anti-phishing), while over half (6) also sought advanced education for security experts.
    
    %just rewrote to reduce words used
    %In contrast, most participants from \textbf{SMEs} (8 of the 11) stated interest in courses for preparing for the risk management capabilities for meeting requirements by the new EU laws. Also, most of the participants from SMEs (8 of 11) indicated a need for basic cybersecurity education, including anti-phishing training, and more than half (6) were also interested in advanced education for security experts.
    %
    \item Both \textbf{public sector organisations} indicated high interest in risk management courses for preparing for the new laws, and stated a high need for basic cybersecurity courses, including anti-phishing training for lay users. One of them also indicated a high interest in advanced cybersecurity courses for experts.
\end{itemize}
About 85\% of the respondents were interested in cybersecurity training and education (which may, however, not be surprising, as probably primarily those who were interested took the time to respond to the survey). 12 of the 19 respondents chose a series of shorter (2--3 hours) lessons as a preferred length, and one-day training events were also chosen as suitable options by half of the participants. 
The open-ended questions showed that, beyond courses on pentesting, awareness and anti-phishing training or NIS2, also other topics were of high interest, including cloud security, cybersecurity for embedded systems, secure coding, cybersecurity related to AI, IoT and OT, supply chain security, and cybersecurity design principles.

%% file: src/results-interviews.tex
\subsection{Interviews results}
\label{ss-int-results}
While the survey revealed first insights and patterns, more explorative research was conducted via our interviews to gain further in-depth insights and to reason about the current situation, challenges, interests, and needs.
This section presents 
%the themes (printed in \textit{\textbf{italic-bold}}) and 
the results answering our three research questions related to these themes (printed in \textit{\textbf{italic-bold}}) that were derived from the interview transcripts' coding via thematic analysis.

\subsubsection{Results related to RQ1.}
%: What is the organisation’s situation and readiness regarding Cybersecurity?
At the start of the interviews, participants were asked about the organisation's current situation regarding cybersecurity.
The following themes and results emerged:
\paragraph{\textbf{Cybersecurity threats, human factors and security awareness:}}
Participants reported a broad range of threats that their organisations are exposed to, also due to an increased sophistication of attacks (P2), conducted by both insiders and outsiders. 
All participants reported that cybersecurity threats related to human factors play an important role. There is an increase in the number and sophistication of social engineering and phishing attacks, even though awareness and practical knowledge regarding phishing would increase (P1). 
P2 describes his experiences with simulated phishing experiments: ``\textit{bank account phishing is very rare in that uh, some that people fall for}\ldots\textit{the most clicked simulation is one that um, uh, was sent out about free iPhones from the IT department}''. Also, misconfigurations cause security problems (P1). 

Short-staffing of qualified personnel with cybersecurity expertise and the lack of cybersecurity knowledge or interest were mainly mentioned for the public sector (P1) and the IT services company (P4) regarding a lack of security knowledge and awareness of their customers.  

While P2 thinks that recent cybersecurity incidents worldwide, and specifically the ransomware attacks that affected V\"{a}rmland against Coop and Tietoevry (see Section \ref{s-intro}) have raised security awareness of its users, P4 observed no significant impact on the (lack of) security awareness of their customers, pointing out that they lack awareness of security risks: ``\textit{The manager and the CEO usually say `yes, but nobody is interested in us.' We sell flower pots, or we bend sheet metal. Or we service cars or whatever it might be. It doesn't really matter. It's the same across the board, so `nobody's interested in us.'\,''} 

The recent substantial investments in cybersecurity and in the defence industry in Värmland may also further increase the security awareness within IT companies in Värmland that deliver these industries, as P4 expects.

\paragraph{\textbf{Organisational security measures:}}
For analysing the cybersecurity situation, needs, and educational measures for those organisations that develop security products or services, it is important to differentiate between organisational cybersecurity and product security, as especially highlighted by P3, P5, and P6. Therefore, we first summarise the findings related to organisational cybersecurity measures and situations in this and the next theme. Aspects related to product security are addressed later in the paper.

P1, who works in the public sector, reported that they use the method support from MSB~\cite{msb2021metodstod}
%The participant from the public sector (P1) reported that they use the method support from MSB~\cite{msb2021metodstod}
%\footnote{MSB, Metodstöd för informationssäkerhetsarbete, https://www.msb.se/sv/amnesomraden/informationssakerhet-cybersakerhet-och-sakra-kommunikationer/arbeta-systematiskt-informationssakerhet-och-cybersakerhet/metodstod-for-informationssakerhetsarbete/} 
for systematically working with information security based on the standards for information security management systems based on the ISO/IEC 27000 standard series~\cite{iso27001-2022}. The other organisations work systematically with information security by following the ISO 27000 standards for conducting risk analyses and implementing and following up security measures. 

%Implemented security measures that were explicitly highlighted included information classification on documents/information security classification (P4, P5), security certifications for employees (P6), access control management (P4), multi-factor authentication (P2, P4), security patch updates (P2), firewalls (P2), VPNs (P2), network segmentation, SOC (security operation centre) and SIEM (security information and event management) that were used (P4), regular external pentesting as a service (P2) and external security audits (P3, P5, P6) with monthly reporting (P3),  accreditations of produced software/products, physical protection, organisational measures and compartmentalisation (P5, P6).
%P1 mentions the challenge of keeping security systems and components updated and the need for an agile security patch system, and the high efforts and time needed to keep their secure systems updated.

\paragraph{\textbf{Security readiness level:}}
Organisations typically follow the ISO 27000 standards for risk analysis and measuring security readiness.
Cybersecurity readiness for the public municipality sector is also typically measured by participating in MSB's cybersecurity check (``cybersäkerhetskollen'') \cite{msb2024cybersakerhetskollen}.
%\footnote{MSB, Cybersäkerhetskollen 2024: Det systematiska cybersäkerhetsarbetet i den offentliga förvaltningen, https://www.msb.se/sv/publikationer/resultatredovisning-av-cybersakerhetskollen-2024--det-systematiska-cybersakerhetsarbetet-i-den-offentliga-forvaltningen/}. 
P1 reported that MSB's cybersecurity check shows that at least the basic cybersecurity level (level 1) is achieved. As the latest MSB report~\cite{msb2024cybersakerhetskollen} also showed, the systematic information security work in Sweden's municipalities is still considerably low and subject to improvement. P1 also suggests that security education can help to that end: \textit{``I think we all landed on the lowest level except for one part, and that was the school\ldots they were one level above because they have developed their training program for even the students and the employees.''}
Managing and harmonising cybersecurity work across different sub-organisations constitutes, however, a challenge, especially for a municipality, which comprises sub-organisations and services (including, for instance, schools, elderly care, cultural and social services, leisure facilities, energy and water supply) with different cybersecurity challenges and requirements, as P1 points out.

The higher education institute (P2) reports that while they do not have a systematic approach for measuring security readiness, they can assess the cybersecurity readiness level via the number and kind of security-related IT service requests/errands that they receive from employees or students and by the number of employees participate in online training sessions using a third-party cybersecurity awareness and education tool. Especially the number of users that ``click'' and fall for simulated phishing attacks issued by that tool provides insightful indications: \textit{``we have some statistics\ldots and we can see what type of simulations they [our users] fall for.''} Apart from that, they have no systematic way of measuring security readiness at the moment, even though, according to P2, it would be a good idea to have.

While P4 %and P3
states that security standards for their companies are high, they report a lower security readiness level for their clients.
%and at the product development side
%.
P5 and P6 report a high security readiness level for both their organisations and for their product security development. However, as P6 points out, the level of product security and product security development is more difficult and not straightforward to define and determine, in contrast to the enterprise security level, which can be measured via security certification and external audit services.
 
\paragraph{\textbf{Addressing product security requirements:}}
The four companies that develop products and services discussed the need for different security measures to enhance product security and/or increase security awareness for developers and customers. P3 points out the need for enhanced cybersecurity for embedded systems and on the physical product level:
``\textit{%I've been in the IT business for 30 years, and, uh,
I think for 20 years, it was a lot of focus about, uh, yeah, you could connect to a network and you can hack into to several, uh, computers that are connected to the network. Today we are on the same page for the embedded system because if you take a car or some else some other product, you have several nodes in the system, several computers or ECUs, and uh, you have uh, you could have the Ethernet network or you can add a CAN network and so on. But there are no encryption about the traffic or the communication between the nodes.}''

New legal requirements pursuant to the CRA regarding  security patch updates pose challenges for product development, especially for IoT and embedded systems:
``\textit{New here is that when you make a product and it leaves a factory, then you are\ldots responsible that for that time you leave a factory, you comply with all the regulations. But no, you have to comply. Always. Lifetime? Yeah\ldots If there are any flaws in the security, then you have to patch that security\ldots over lifetime}'' (P3). Therefore, an agile security patch system and special training for handling security patches and ``\textit{compliance for lifetime}'' (P3) is needed.

Product and software accreditation and lifecycle security assurance for products, including design, implementation, and operation phases, is important to guarantee products with high security requirements, as P5 explores.

P6 discusses challenges for embedding security practices at all parts of the product development lifecycle. Right from pentesting ``\textit{through to, you know, continuous integration, continuous development, that whole cycle and how it fits in at every part of that into when it's operational out in the field}'' (P6).
   
\paragraph{\textbf{Lack of awareness of security as a non-functional requirement: }}    
P3 and P4, i.e. the representatives of the two companies that produce products and services for markets outside the security and defence sector, discuss the lack of security knowledge and awareness of their software developers and/or customers regarding the importance of addressing security as a non-functional requirement.

P3 argues that software engineers typically tend to focus on the main functionality of a product, while paying less attention than needed to security as a non-functional requirement. \textit{``It's knowledge. It's, uh, you're afraid of something new, and it's not, it's not building functionality. You want to see progress in building functionality. Cybersecurity and testing is. It's almost backwards\ldots Yeah. Non-functional\ldots the threshold for, being interested in cybersecurity---it's too high. So that's why I will put all this, for our software engineers to our focus.''}

P4 emphasises a lack of awareness of their customers of the legal requirements for cybersecurity, i.e. the need to consider security as a non-functional requirement, also for the reason of achieving legal compliance: \textit{``it's the tenth percentile only who have understood this so far...that there will be legislation even now.''} 

\paragraph{\textbf{Impact of cybersecurity regulation on educational needs:}}
The wave of new cybersecurity laws %, including GDPR, NIS2, CRA, and DORA, 
that organisations need to comply with, as well as their impact on the organisations and their educational needs, play an important role and are discussed by most participants. 

The requirement of the NIS2 Directive for mandatory cybersecurity training for organisations working with critical information infrastructures was mentioned as a benefit for motivating organisations to invest and provide cybersecurity training for their employees, which especially P1 highlights as an argument that they can bring forward to the board of his municipality (``kommun styrelsen''): 
 ``\textit{I think it does help being, uh, how do you say kommun styrylsen as the, um, uh, group that is that they pinpointed as the group that they intend they need to have get education within those nine risk fields. Mhm. But that also they say, I mean in that suggestion or position that all employees should be given the opportunity to receive education. So I mean now it's the law\ldots But I think it makes it easier to actually enforce or hopefully budget as well for it.''}

 P3 discusses future challenges for complying with the CRA's requirements for supporting life-long security patch updates for their products. Additionally, P3 also mentions new laws that put requirements on the automotive and mining industries, and autonomous mobile solutions that are impacting their business, and the required competence profiles of their employees.

A general unwillingness by customers to invest in cybersecurity and an unawareness or lack of interest in legal compliance for cybersecurity is a problem experienced by P4, as aforementioned. For increasing awareness, they often ask customers to sign a statement that, even if it is recommended for achieving legal compliance, they choose not to invest in a higher security level, e.g. that they do not want to invest in multi-factor authentication (even though it is required by NIS2 for critical sectors and the Swedish Data Protection Agency, as a rule of thumb, for complying with the GDPR when processing sensitive information). Asking the customers to take over this responsibility and liability for their choices to save money at the expense of having less security and legal non-compliance usually makes customers reconsider: 
%Now we've talked about two-factor authentication. We'll introduce that for you. “No”, says the customer, ”we won't do that. It's cumbersome”. It costs nothing extra at all. It's part of the service,” I say. “Yes, but it costs to start, right?” Yes, it does. “No, we don't want that”, says the customer.
%
``\textit{Yes. Sign here. They don't want.  Yes [laughter]\ldots an anti-contract\ldots so that's why you should sign this. Because then you're telling me here that you've understood and you're abstaining because otherwise something will happen soon, and then you'll come to me and say, `you should have said that!' you have a piece of paper that says I did. And it's mostly to get the customer, think outside the box}.''
    
\paragraph{\textbf{Benefits of education:}}
In addition to facilitating legal compliance, participants acknowledge various other benefits of cybersecurity education, even though P5 notes that such benefits may often rather be long-term. 
Short-term benefits are, however, apparent regarding the organisation's vulnerability to phishing attacks. While the percentages of employees who were falling for (simulated) phishing attacks had been increasing, this trend stopped in the last year when the organisation started with anti-phishing training for its employees, as P2 reports. 
%"But the most clicked simulation is one that um, uh, was sent out about free iPhones from the IT department", while "bank account phishing is very rare in that uh, some that people fall for".
P2 measured that the anti-phishing training, provided by their third-party cybersecurity awareness and education tool, decreased the percentages of users falling for simulated phishing attacks from 5.6\% to 3.5\%. ``\textit{So there's more awareness now with this training also}'' (P2). 

P3 also highlighted the potential benefit of raising awareness for security as a non-functional requirement, while P5 pointed out that education is also about training employees to think and reflect, which is important for an organisation: ``\textit{You also educate how to think. Yes. To take new knowledge in your head or to have this discussion in your own head and with other people, of course. So that is also very important}.''

Education of software engineers on product development security is also seen as a necessity for coping with customers' demands, as P3 mentions: ``\textit{Just a game changer here for for a company like us, if we don't train the personnel and don't cope with the demands from our customers, then we don't have any services to deliver, uh, in maybe in five years. So, uh, that's very important for us}.''

Receiving certifications for attended security training is also discussed as important or required for employees of companies in the security and defence sector (P5, P6), but also for the security consulting business (P4), concerning their suppliers, customers, and when hiring new employees to prove competence.

\paragraph{\textbf{Existing education:}}
All participants' organisations offer ``onboarding'' cybersecurity training for their new employees, either in the form of online lectures and exercises that need to be passed, or through introductory lectures. However, as P2 reports for their higher education institute, participation in these courses has been voluntary, and only 60\% participated in the online courses.
    
Moreover, the organisations typically have online anti-phishing and awareness training by external providers with simulated attack exercises. 
Beyond such online training on security awareness, especially the organisations in the public and educational sector are currently lacking continuous education and training on organisational security for their employees, as P1 and P2 point out, with the exception of security education for IT personnel at annual planning days (P2). P2 also mentions that cybersecurity education is, in general, only offered to employees, but not to students.

In addition to education on organisational cybersecurity, the companies offer different training courses for product and product development security to their software engineers. External consultants usually give courses on penetration testing or related topics, as stated by all participants from product development companies (P3, P4, P5, P6).
      
P3 states that their company has the strategy to spend 4\% of its profit on employee training, including cybersecurity and machine learning courses. Cybersecurity courses related to their products can partly be provided by their products' customers who have the required domain knowledge, as P3 reports.

\subsubsection{Results related to RQ2.} 
In the second part of the interviews, participants were asked about their cybersecurity education and competence development challenges. Derived themes and related results are:

\paragraph{\textbf{Recruitment of professionals:}}
Cybersecurity competence of employees can usually be directly obtained by recruiting cybersecurity experts with the required knowledge and skills. However, the challenge of recruiting cybersecurity experts due to a security skills gap is experienced by all participants. 

The need for cybersecurity and defence sector employees to have security clearance and/or to be Swedish or a NATO state citizen further limits the number of suitable candidates, as P5 and P6 explain. At the same time, the demand for hiring cybersecurity experts in the defence sector has increased significantly in recent years due to the recent massive investments in cybersecurity and defence products in Sweden, as mentioned by P5 and P6.
    
%    \item internal training absent
%   \item absent organisation security strategy 
\paragraph{\textbf{Challenges and obstacles for cybersecurity education:}}
 The participants experienced various challenges in providing cybersecurity education within their organisation.
 
As aforementioned, P1 discussed his challenge as a CISO of a municipality with a broad range of sectors and activities to arrange cybersecurity education due to the diversity of users in the different sectors, with different tasks, who have different educational demands. P2 discusses the challenge for cybersecurity education to keep up with the advances of the cybersecurity field and the changing security threat landscape. P5 mentions the general problem that education is often cut first if the workload is high, and still be able to meet deadlines.

Nonetheless, many challenges and obstacles that participants highlight relate primarily to providing cybersecurity courses for secure product development rather than to providing courses on organisational cybersecurity.

The problem of courses on product development security being less available on the market than courses on enterprise/organisational security was raised by several participants. This becomes even more of an issue if software developers must be trained in specific technical domains or application areas. For instance, P3 discusses the lack of courses on embedded systems security, even though this topic is increasingly becoming important, also for complying with laws, such as the CRA, as discussed above. When they looked for respective course offerings, they could only find one course in Europe matching their needs: ``\textit{I've been searching for companies that could give us training in cybersecurity for, I will say, embedded system or a product cybersecurity. Uh, and um, we've done I've done such, uh, just search around and find one company in Austria}'' (P4). 
For developing security for embedded systems, specific competences are needed, related, e.g., to lightweight and secure encryption, hardware security, Linux system hardening, regulations to be applied, how to achieve legal compliance, and also to the application domain, such as connected vehicles or autonomous systems. Hence, newly designing and holding such a course requires various expertise and is not easily done. Even with the increase of smart devices and IoT and the need for cybersecurity at the physical product level, the specific topic of embedded systems security was not part of the cybersecurity curricula at Swedish universities, as P3 observed and criticised.

Also, P6 emphasises that designing courses on product development security poses challenges, as it needs to include teaching on secure software development, including pen testing, related to operational security of the specific application area, but ``\textit{there's very few people who really maybe know that area inside out.}''
Moreover, P6 brings up that according to their experiences, academia cannot easily provide contract teaching to meet industrial needs, as it would lack the required flexibility. Academics are usually tied up time-wise and thus cannot easily put aside the time and resources needed for course development on shorter notice, as experienced by P6.

\subsubsection{Results related RQ3.} Themes and related results derived from discussing interests and requirements regarding cybersecurity education are the following: 
%What are their requirements regarding cybersecurity education (regarding topics, target groups and course forms)? 

\paragraph{\textbf{Regulatory requirements for education:}}
As discussed above, the new cybersecurity laws and especially NIS2, put requirements for systematic cybersecurity work and cybersecurity education and training within the organisations working in the critical and important sectors listed in NIS2. Especially, P1 referred to these NIS2 requirements as motivation and arguments for further investments in cybersecurity education and work within the municipality.

Customer education on the product's security and how to use it securely is required during the accreditation phase, according to the Swedish Defence Forces security requirements for IT systems ``krav på IT-säkerhetsförmågor hos IT-system'' (KSF) \cite{ksf3.1}, 
%~\footnote{https://www.fmv.se/globalassets/dokument/om-fmv/sakerhetsskydd/ksf-3.1.pdf}
as P5 explains.
\paragraph{\textbf{Cybersecurity topics of interest and need for competence development:}}
As discussed above, especially P3 and P6 highlighted the interests of their organisations in cybersecurity courses for software developers on secure product development, linked to the specific operational environments of the products, even though, as P6 also pointed out, the needs and requirements for product security may be more challenging to define.

Furthermore specific cybersecurity topics, where the participants saw an interest or need for competence development of their employees, customers or users included: Risk management (P2), cybersecurity regulations and legal and GDPR compliance (P1, P2) also for customers/managers (P4), AI competence for management and AI driving licence (P1), organisational rules for app (TikTok) and rules for AI use (ChatGPT) for teachers (P1), advances phishing training (P1), Azure/MS365 security, host hardening (P3, P4), cloud security and secure cloud configuration (P2), ethical hacking (P2) and pentesting, also with a focus on embedded systems (P3), and
%(even though that many organisations (e.g. of P1) have also outsourced pentesting education),  
cryptography and specifically practical and secure message/light-weight encryption (P3).

P2 states that the most important action would be cybersecurity education for their IT support staff with an IT background.

\paragraph{\textbf{Preferred instance and forms of cybersecurity education and training:}}
The need for continuous education is discussed by most participants. P1 mentions the need to complement the onboarding training for new employees with continuous training, e.g., training related to social engineering or ensuring legal compliance in practice: ``\textit{and that's what we need to address to get continuity in training\ldots I mean we can't expect you to, to remember what you did if you've been employed for seven years or how is it possible to even.}''

For practical reasons, shorter online presentations and training sessions on a higher level are preferred for employees, who can be part of a longer education scattered over time and/or given continuously (P1, P2, P3, P5). 

It may be a challenge to conduct education on the users' devices, as they can be classified: ``\textit{but the problem is that resources are often used for simulations or for testing or so we can't perhaps do education on it in a very easy way}'' (P5).

Benefits of on-site training, which also enables group discussions among employees (P5), are pointed out. On-site and longer training workshops (e.g., 1-2 days) for technical personnel and system admins (P1, P2, P3), e.g., on ethical hacking or pentesting, or for customers if combined with a social event (P4) were mentioned as examples.
%P3 also regularly offers 2-hours inspiration workshops introducing into a new area (e.g. new regulation) and regular training sessions, e.g. personality traing program of 4 hours.
P2 suggests that cybersecurity training within their organisation should be compulsory to get all employees involved. 

\paragraph{\textbf{Information exchange network/communities on cybersecurity:}}
The participants P5 and P6 from the large companies bring up the usefulness of discussion networks consisting of security experts from other companies and organisations, including from academia, with similar cybersecurity needs and interests. 
According to P5, who used to be the chief digitalisation officer for a large company in Värmland before his current assignment, participating in a digitalisation community network with several other larger companies, where cybersecurity was discussed to some extent, is of value. 
P5 perceives this exchange of knowledge and experiences on a higher level as very useful and therefore would appreciate a similar community network in Värmland exclusively for cybersecurity:
``\textit{If we could build a similar joint venture in IT security on the critical infrastructure\ldots I would like to have perhaps more collaborative discussion networks in this region between companies. I think that it could be, and I think university is like the spider in the middle. But the companies could also discuss with each other because there's a, uh, the knowledge is in the whole community.}''
%Yeah, it could be like that. Yeah. Uh, and also the university has a place for discussions because we need the students. We could meet, the students. We could have a discussion with bachelor thesis up to research level or whatever.}" (P5).....

P6's employer is part of an expert cybersecurity learning lab with other companies and Swedish universities, where participants can discuss experiences and measures regarding product security education: ``\textit{it is the competence there\ldots and which is one of the reasons, and also why we're working across industry in this \ldots consortium as well. Because\ldots there is a lot of value in getting experts to talk to each other\ldots what we're actually trying to do within [this consortium] is getting experts to network from across companies. Um, and maybe that academia is playing a facilitated facilitation role in that\ldots That we see is very important.}'' Nonetheless, P5 and P6 also note the challenge of discussing security-classified topics, which requires keeping the discussions on a higher level.

%% file: src/discussion.tex
\section{Discussion}
\label{s-discussion}
%
%of main results, related work, ways ahead, limitations....
%Some key findings from our interviews can be summarised as follows:
Our study provided several findings that are of relevance for understanding the cybersecurity readiness, specifically in terms of competence levels, challenges, and needs for cybersecurity education, that may be partly specific to organisations in Värmland or generally in Sweden and may also partly hold for other regions in the world.

\subsection{Insights from key findings from the interviews}
%The results from the online survey show that most large companies in Värmland were less interested in basic courses for organisational security including anti-phishing and awareness training, which suggest that they probably have already such courses in place, and they also seem to be better prepared for meeting risk management requirements of the cybersecurity laws. They are therefore also rather interested in courses for training IT security specialists, while SMEs and the public sector are interested in a broader range of new cybersecurity education for their employees and/or users. 
Key findings from our interviews provided us with important insights, which can be summarised as follows:

\begin{itemize}
    \item Organisations typically have at least introductory (``onboarding'') cybersecurity education for new employees and additionally shorter online training, especially related to social engineering and phishing, involving simulated phishing attacks, in place. However, other types of shorter online cybersecurity training and continuous cybersecurity education are lacking for the public and educational sector organisations that were interviewed, even though there is a recognised need (also for complying with NIS2), interest and intention for implementing a broader continuous education for the users and employees' competence development for the future.

    \item It is important to distinguish demands for cybersecurity and for cybersecurity education for organisational security from demands for product security and education for secure products and their development.
    For addressing organisational security requirements, measures are well defined, and there is a broader range of available courses and certification schemes for organisational and enterprise security. For product security, especially in relation to specific operational environments, the demands are more complex to define, and suitable education (e.g., on embedded systems security) is hard to find.

    \item Security as a non-functional requirement is often poorly understood by software developers and customers of products and services outside the security and defence industry. Thus, cybersecurity education and training of software developers and customers in these companies is important for raising their cybersecurity knowledge and awareness.

    \item The new European cybersecurity regulations are putting requirements for implementing cybersecurity education in organisations. They point out the need for organisations to invest in cybersecurity education and to increase their expertise and cybersecurity strategies for achieving legal compliance.

    \item The large-sized companies that were interviewed appreciate regional or national networks with other companies and academic partners, for sharing experiences and exchanging ideas with peers on a high level on how to approach these challenges, and/or how to design and implement education for secure product development.

    \item The cybersecurity skills gap and diverse cybersecurity competence expertise requirements needed by the different organisations and in different operational environments are well noticeable. For a municipality in Värmland, security competences need to be broad due to its wide range of different types of sectors and activities, while companies need security competences for various types of IT systems, services, and operational environments. Graduates from universities in Sweden will not necessarily have all these competences required by the different types of organisations and their application and operational environment, as also e.g. P4 noted related to the specific competences needed for embedded systems security and in line with what P6 discussed related to the specialised competence required for product security in specific operational environments.
 
\end{itemize}

%Discussion: What findings are typical for Sweden/Värmland, Europe - and which are general findings?
\subsection{Overcoming the cybersecurity skills gap}
Findings related to the cybersecurity skills gap are not directly specific to Värmland and have been observed for several years as a problem in Europe and worldwide. The latest ISC2 report shows a worldwide increasing cybersecurity workforce gap 
%with more than 390.000 cybersecurity experts missing in Europe 
and reports that 90\% of organisations have a skills gap within their security teams~\cite{isc2:2024}. Nonetheless, the presence of the defence industry in Värmland, combined with recent higher investments and demands for cybersecurity and defence products, have made the skills gap even more severe and noticeable in Värmland.
%Moreover, while Sweden ranks as one of the most advanced countries regarding digitalisation in the world, it has fallen behind in cybersecurity readiness, where it takes only the 14th position. 
For this reason, increasing efforts in cybersecurity education for organisations in Värmland and Sweden becomes even more important. 

A recent study~\cite{Behzadi1875328} analyses how well the cybersecurity educational programmes at Swedish Universities can prepare students for diverse cybersecurity roles identified by ENISA's European Cybersecurity Skills Framework ECSF~\cite{ENISA2022ECSF}. It shows that none of the Swedish Universities offers courses that prepare students for every role in the ECSF framework. This confirms the observation that graduates from universities in Sweden may not have all the required competencies for different types of jobs and organisations, and they will still need cybersecurity education and training at their jobs. Nonetheless, the same study shows that every role specified by ECSF is addressed by at least one course at Swedish universities.
%, but especially specialized courses, such as for instance on 'Cybersecurity for Artificial Intelligence (AI)', were limited to only one university~\cite{Behzadi1875328}. 
Hence, cooperation and sharing of courses and courseware between Swedish and Inner Scandinavian universities is key to meeting the required competence skills for different cybersecurity roles in different organisations.

\subsection{Industry-academia cybersecurity expert networks}
For bridging the cybersecurity skills gap, the EU Commission recently launched an Industry-Academia Network to ``\textit{strengthen the links between industry and academia to boost cybersecurity skills, ensuring the European cyber workforce is adequately equipped to meet growing demands in the sector}.''\,\footnote{\url{https://digital-strategy.ec.europa.eu/en/news/commission-launches-industry-academia-network-bridge-cybersecurity-skills-gap}}
Our interviews also revealed that the large companies appreciate such type of industry-academia networks with other industrial and academic partners from Värmland, Inner Scandinavia, and/or Sweden.
Such industry-academic networks can help to exchange insight and knowledge on topics of high relevance, such as what educational, organisational, and technical means are needed to comply with new regulations or enforce trustworthy and robust AI. Additionally, such networks were also considered important for discussing and addressing challenges relating to the training of software and system developers in secure product development.

Sweden and other Nordic countries are known for their flat hierarchies in organisations and their consensus culture~\cite {hofstede, meyer2016culture}. In Sweden, the consensus and ``fika paus'' (coffee breaks with colleagues) culture, which is dominant at workplaces, supports informal discussions, team-building, and the decision-making process, and leads to a better collaboration among colleagues~\cite {meyer2016culture, ta2017cross}. 
Since Swedes are traditionally used to and value collaboration at work, it may explain and motivate why Swedish company representatives value collaboration networks, as indicated in our interviews. 
Hence, supporting organisations in Sweden and Värmland to establish and/or maintain such cybersecurity expert networks for ensuring that organisations are well equipped to meet growing demands for cybersecurity, similar to  ENISA's approach, may be a good investment to address cybersecurity skills gap challenges. Moreover, to utilise the competence of the Norwegian cybersecurity hub in Innlandet, establishing a cross-border industry-academia cybersecurity expert network offering regular discussion rounds and workshops for competence and experience exchange should be further considered.

\subsection{Limitations}
Our interview study included six participants, which may be considered a limitation; however, they were key stakeholders who are typically difficult to access. Furthermore, the aim was to conduct explorative research and to derive qualitative (rather than quantitative) results from relevant sectors. %was based on limited number of participants from organisations that participated in our survey (N=19) and interviews (N=6). Therefore, 
While there is the risk that the results may mainly reflect the views from the participating organisations, 
%Still, the survey allowed us to derive common patterns and 
the interviews revealed recurring codes/themes across public sector bodies, SMEs, large companies, and product/service providers.%for different types of organisations from the public sector, SMEs, large companies, and companies developing different types of products or services. 
This indicates that code saturation was largely reached, i.e. hardly any additional aspects were identified, and further interviews would likely yield redundant information. Code saturation is a recognised criterion for determining when to stop data collection, supporting the sufficiency of our six interviews.

%This means that code saturation was largely reached, i.e. we reached a point where hardly any additional aspects were identified and where the codebook began to stabilise. Code saturation is an accepted concept in determining when to stop collecting and analysing data, suggesting that, in our case, further data collection via additional interviews would likely yield redundant information and that six interviews could therefore be regarded as sufficient.

%% file: src/conclusions.tex
\section{Conclusions}
\label{s-conclusions}
Our study provides several findings and important insights regarding the cybersecurity situation at different types of organisations in Värmland, the challenges for delivering cybersecurity education, and their needs and interests for cybersecurity and education. Our findings suggest in particular to invest more efforts in developing continuous cybersecurity training especially for the public sector, creating more specialised courses for secure product development (e.g., for embedded systems or specific operational environments), new course offerings for preparing for the requirements posed by new European cybersecurity-related laws, and the exchange of access to courses/courseware among Swedish universities and across borders, and maintain and establish industry-academia cybersecurity expert networks for exchanging ideas, insights, expertise in the region, in Sweden and across borders. Such recommendations may also hold for other regions or countries with similar public sector structures, emerging cybersecurity challenges, and a need for specialized training, suggesting that while some findings may be context-specific, many of the recommendations could be applicable more broadly.
%future work:
The results of this study and the study conducted in parallel in Innlandet Norway will allow us, as part of our future work, to analyse similarities and differences between organisations in Värmland and Innlandet as well as the potential for strengthening cybersecurity in Inner Scandinavia in cooperation through addressing educational needs in the region.

%% file: src/appendices.tex
\appendix\ %Note that paper can be 16 pages without ref and appendix and 20 pages in total. 

\section{Interview Guide}\label{interviewguide}

The interview script is presented in this section. All questions were optional and could be left out depending on the flow of the conversation.\\

\noindent\textbf{Introductory Questions}
\begin{enumerate}
    \item Could you tell us about your responsibilities/interests for training and education?
\end{enumerate}

\noindent\textbf{Status of the Organisation}

\noindent\textit{Risks}
\begin{enumerate}[resume]
    \item What are the most significant cybersecurity threats for your organisation?  
    \textit{Probe: Which of the following threats do you consider in your organisation: outsider (hacker) attacks, including malware, social engineering attacks, insider attacks, human errors and misconfigurations, etc.}
    \item How are you addressing these risks?
\end{enumerate}

\noindent\textit{Current Situation}
\begin{enumerate}[resume]
    \item How is the level of cybersecurity expertise and competence of the users in your organisation?
    \item How would you describe your organisation’s cybersecurity readiness level?
    \item How does your organisation assess and update its cybersecurity risk profile?
    \item What existing measures does your organisation have for cybersecurity training and education?
    \item What cybersecurity competence development offers are currently used or exist in your organisation?
    \item Can you describe a situation where cybersecurity training directly benefited your organisation?
    \item What are the efforts for the recruitment of cybersecurity experts for your organisation?
\end{enumerate}

\noindent\textit{Challenges}
\begin{enumerate}[resume]
    \item What challenges do you foresee when it comes to cybersecurity training and education?  
    \textit{Probe: What challenges do you face in creating awareness and recruitment of cybersecurity experts?}
    \item How would you estimate the relation between efforts/costs and benefits of cybersecurity training and education?
    \item What are barriers to involving employees in cybersecurity training/education?
\end{enumerate}

\noindent\textbf{Requirements / Concrete Needs for Educational Offers}
\begin{enumerate}[resume]
    \item Which of the following is most important and why? Please motivate your previous answers:  
    \begin{itemize}
        \item Advanced education of cybersecurity experts (e.g., penetration testing)
        \item Cybersecurity education of employees with an IT background (e.g., introduction to penetration testing)
        \item Cybersecurity education for lay users (with no IT or technical background), e.g., through a basic cybersecurity education including anti-phishing training
        \item Courses on risk management for legal compliance (GDPR, AI Act, Cyber Resilience Act, NIS 2, etc.)
        \item Further specific topics?
    \end{itemize}
    \item Is customisation of the training content to the needs of your organisation important?
    \item Preferred length of courses:
    \begin{itemize}
        \item 1-day training
        \item 2--3 days training
        \item Series of shorter (1--3 hours) lessons
        \item Others?
    \end{itemize}
    \item Preferred locations:
    \begin{itemize}
        \item On site -- at X university (redacted to fulfil anonymity requirement)
        \item On site -- at your company
        \item Online (live) teaching
        \item Recorded videos and quizzes
    \end{itemize}
    \item What access to (internal or external) resources and infrastructure (e.g., cyber range, hacking tools) is needed for education/training?
    \item How frequently should such educational offers be given and updated?
    \item Are you using or are you interested in cybersecurity contract teaching?
\end{enumerate}

\noindent\textbf{Need for Private/Public Partnerships}
\begin{enumerate}[resume]
    \item What is your position on collaborating with other private/public partners?
\end{enumerate}

\section{Survey Instrument}\label{survey}

\noindent\textbf{Survey on Cybersecurity Educational Needs.}
(Reproduced from the survey instrument that participants answered. Note that answer options to the questions are listed after each question, separated by comma.)%

\begin{enumerate}[label=\textbf{Q\arabic*:}, leftmargin=*]

\item How do you identify yourself? Male, Female, Others, Prefer not to say.

\item Name of your organisation (optional, free text).

\item Please specify your organisation type: Micro Company ($<10$ employees), Small or Medium-size Enterprise/Industry (SME, $\leq 250$ employees), Large Enterprise/Industry ($>250$ employees), Public Sector (authority), Prefer not to say, Other (please specify). 

\item What is your current role within your organisation? Executive/Management, IT Security Administrator/Specialist, Other Type of IT Professional, Non-IT Professional, Other (please specify). 

\item Is your organisation interested in cybersecurity education and competence development for employees? Yes, No, Do not know. 

\item What cybersecurity training or education is needed in your organisation (beyond what is already offered)? (free text)

\item Is there a need for advanced education of cybersecurity experts (e.g., advanced testing/penetration testing or Capture the Flag (CTF) exercises)? Five-point Likert Scale answers from \textit{Extremely needed} to \textit{Not at all needed}.

\item Is there a need for cybersecurity education of employees with an IT background (e.g., introduction to penetration testing and CTF exercises)?
Five-point Likert Scale answers from \textit{Extremely needed} to \textit{Not at all needed}.

\item Is there a need for cybersecurity education for lay users (no IT/technical background), e.g., basic cybersecurity, including anti-phishing training? Five-point Likert Scale answers from \textit{Extremely needed} to \textit{Not at all needed}.

\item Are there other cybersecurity courses/topics of interest? (e.g., GDPR, cloud security, forensics, …) (free text)

\item Are there other target groups of interest (beyond employees, e.g., school students)? (free text)

\item What form of training/education would suit your organisation? (select all that apply) 1 day training, 2–3 days training, Series of shorter (1–3 hours) lessons, Others (please specify).

\noindent \textit{What is the preferred length? Please specify (free text).}

\item What is the preferred location? On-site -- at X university [Redacted to fulfil anonymity requirement], On-site -- at the premises of your organisation, Online (live) teaching, Recorded videos and quizzes (asynchronous), Other forms (please specify).

\noindent \textit{Please provide details (free text).}

\item How well do you think your organisation is prepared for complying with the risk management requirements of new EU regulations relating to cybersecurity (incl.\ Cyber Resilience Act, NIS 2, AI Act, Digital Services Act)? Five-point Likert Scale answers from \textit{Very well prepared} to \textit{Not at all prepared}. 

\item Is your organisation interested in developing and training approaches for Risk Analysis to comply not only with GDPR Data Protection Impact Assessments but also with risk analysis requirements in other laws (incl.\ Cyber Resilience Act, NIS~2, AI Act, Digital Services Act)? Yes, No, Do not know.

\item General comments: (free text)

\item Are you interested/available for a follow-up talk/interview to discuss how the project could address cybersecurity education needs for your organisation? Yes, No. 

\item Please leave your name, email and/or phone number for contact: (optional, free text)

\end{enumerate}

%\section*{Appendices:}
%Should we append:

%Survey questions / results? LM: questions only

%Interview guide? LM: I think so

%Codebook? LM: not needed IMO

%LM: add Consent form